\documentclass[fleqn,9pt]{article}
\usepackage{amsmath,amssymb,graphicx,authblk,extsizes}
\usepackage{anyfontsize,fleqn}
\usepackage[font=small]{caption}
\usepackage[top=1.5cm, bottom=1.5cm, left=1.5cm, right=1.5cm]{geometry}
\usepackage{multicol,textcomp,xfrac}
\usepackage[superscript,nomove]{cite}

\renewcommand{\arraystretch}{1.5}

\makeatletter
\renewcommand{\maketitle}{\bgroup\setlength{\parindent}{0pt}
\begin{flushleft}
  \textbf{{\fontsize{24}{30}\selectfont \@title}}
  \newline
  \@author
\end{flushleft}\egroup
}
\makeatother

\newenvironment{Figure}
  {\par\medskip\noindent\minipage{\linewidth}}
  {\endminipage\par\medskip}

\title{A CNOT gate between multiphoton qubits encoded in two cavities}

\begin{document}

\author[1,2,$\star,\dag$]{S. Rosenblum}
\author[1,2,$\dag$]{Y.Y. Gao }
\author[1,2]{P. Reinhold}
\author[1,2,+]{C. Wang}
\author[1,2]{C.J. Axline}
\author[1,2]{L. Frunzio}
\author[1,2]{S.M. Girvin}
\author[1,2]{Liang Jiang}
\author[2,3]{M. Mirrahimi}
\author[1,2]{M.H. Devoret}
\author[1,2,$\star$]{R.J. Schoelkopf}
\affil[1]{Department of Applied Physics and Physics, Yale University, New Haven, Connecticut
06520, USA}
\affil[2]{Yale Quantum Institute, Yale University, New Haven, Connecticut 06520, USA}
\affil[3]{QUANTIC team, INRIA de Paris, 2 Rue Simone Iff, 75012 Paris, France}
\affil[+]{Present address: Department of Physics, University of Massachusetts, Amherst, MA 01003,
USA}
\affil[$\dag$]{These authors contributed equally to this work, $^\star$e-mail:
serge.rosenblum@yale.edu; robert.schoelkopf@yale.edu}

\maketitle
\begin{multicols}{2}
%\section*{Abstract}
\textbf{Entangling gates between qubits are a crucial component for performing algorithms in
quantum computers. However, any quantum algorithm must ultimately operate on error-protected
logical qubits encoded in high-dimensional systems. Typically, logical qubits are encoded in
multiple two-level systems, but entangling gates operating on such qubits are highly complex and
have not yet been demonstrated.
Here, we realize a controlled NOT (CNOT) gate between two multiphoton qubits in two microwave
cavities. In this approach, we encode a qubit in the high-dimensional space of a single cavity
mode, rather than in multiple two-level systems. We couple two such encoded qubits together
through a transmon, which is driven by an RF pump to apply the gate within 190 ns. This is two
orders of magnitude shorter than the decoherence time of the transmon, enabling a high-fidelity
gate operation. These results are an important step towards universal algorithms on
error-corrected logical qubits.}

\section*{Introduction}

In traditional approaches to quantum error correction, bits of quantum information are redundantly
encoded in a register of two-level systems\cite{shor_scheme_1995,steane_multiple-particle_1996}.
Over the past years, elements of quantum error correction have been implemented in a variety of
platforms, ranging from nuclear spins\cite{cory_experimental_1998},
photons\cite{pittman_demonstration_2005} and atoms\cite{chiaverini_realization_2004}, to crystal
defects\cite{waldherr_quantum_2014} and superconducting
devices\cite{reed_realization_2012,ofek_extending_2016,kelly_state_2015}. However, for performing
actual algorithms with an error-protected device, it is necessary not only to create and
manipulate separate logical qubits, but also to perform entangling quantum gates between them. To
date, a gate between logical qubits has not yet been demonstrated, in part due to the large number
of operations required for implementing such a gate. For example, in the Steane
code\cite{steane_multiple-particle_1996,nigg_quantum_2014}, which protects against bit and phase
flip errors, a standard logical CNOT gate would consist of seven pairwise CNOT gates between two
seven-qubit registers\cite{nielsen_quantum_2010}. Previous experiments have demonstrated an
effective gate between two-qubit registers that are protected against correlated
dephasing\cite{monz_realization_2009}. In that case, an entangling gate could be implemented using
just a single pairwise CNOT gate between the registers.

We choose to pursue a different strategy by encoding qubits in the higher-dimensional Hilbert
space of a single harmonic oscillator\cite{gottesman_encoding_2001}, or more concretely in
multiphoton states of a microwave cavity
mode\cite{mirrahimi_dynamically_2014,leghtas_hardware-efficient_2013}. This approach has the
advantage of having photon loss as the single dominant error channel, with
photon-number parity as the associated error syndrome.
Codes whose basis states have definite parity, such as the Schr\"odinger cat code
\cite{vlastakis_deterministically_2013} or the binomial kitten code \cite{michael_new_2016}, can
then be used to actively protect quantum information against this
error\cite{sun_tracking_2014,ofek_extending_2016}.
While preparation of an entangled state between two cavities has been performed
before\cite{wang_deterministic_2011,wang_schrodinger_2016}, a quantum gate between two multiphoton
qubits has so far not been demonstrated. In contrast to gates between two-level systems, which can
be coupled by a linear element such as a cavity bus\cite{majer_coupling_2007}, harmonic
oscillators can non-trivially interact only if they are coupled by a nonlinear ancillary element.
However, the requirement for fast interaction between the cavities without inheriting large
undesired nonlinearities and decoherence from the ancilla, presents a challenge to the
cavity-based approach to quantum error correction.

In this work, we address this challenge by coupling the two cavities to an RF-driven ancilla transmon. The cavities interact sequentially with the ancilla to effectively implement a CNOT gate between the two encoded multiphoton qubits.
We generate a high-fidelity multiphoton Bell state, perform quantum process tomography of the
gate, and apply the gate repeatedly in order to quantify imperfections in the operation.
The number of CNOT operations that we can coherently apply is $\sim 10^2$, bringing this
gate within the regime required for practical quantum
operations\cite{nickerson_topological_2013,barends_superconducting_2014}. We also measure the
undesired entangling rate between the cavities during idle times, and infer a high on/off ratio of
the entangling rate\cite{dicarlo_demonstration_2009,bialczak_quantum_2010} of $\sim$ 300. This
figure of merit is important since undesired cross-talk is often a major hurdle when trying to
scale up to a larger number of qubits.

\begin{Figure}
\centering
\includegraphics[scale=1]{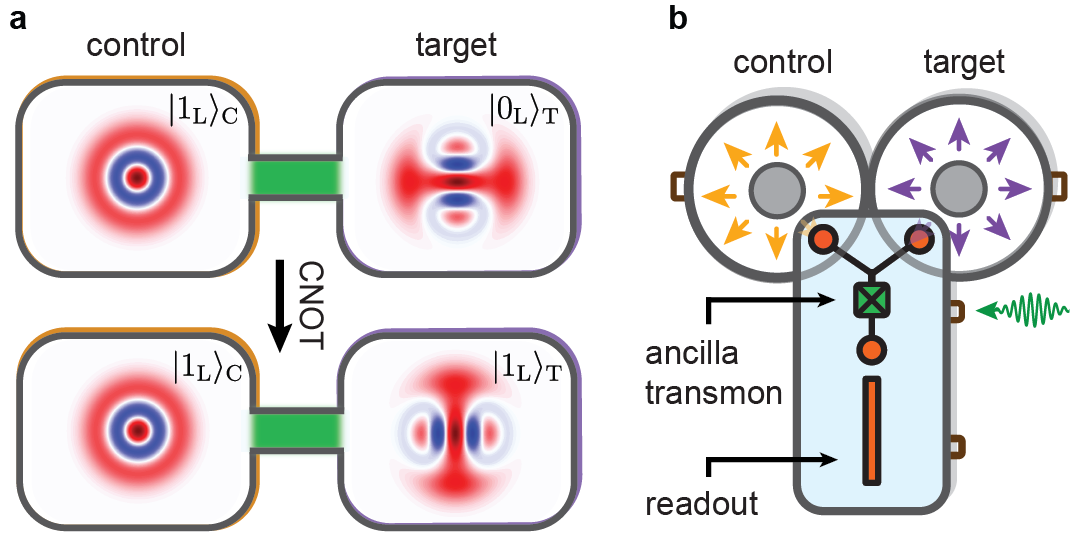}
\captionof{figure}{\textbf{Experimental implementation of an entangling gate between multiphoton
qubits encoded in two cavities. a,} Example of the CNOT operation. In the initial state,
illustrated by the Wigner distributions in the top panel, the control qubit is in
$|1_\textrm{L}\rangle_\textrm{C}$, and the target qubit in $|0_\textrm{L}\rangle_\textrm{T}$ (as
defined in equations (1) and (2)). Under the action of the CNOT gate, enabled by a nonlinear
coupling between the cavities (in green), the target state at the output (bottom panel) is
inverted to $|1_\textrm{L}\rangle_\textrm{T}$.
\textbf{b,} Sketch of the device, which is housed inside an aluminum box, and cooled down to 20
mK. The control and target qubits are encoded in photon states of the fundamental modes (yellow
and purple arrows) of two coaxial cavities with frequencies $\omega_\textrm{C}/2\pi=4.22$ GHz and
$\omega_\textrm{T}/2\pi=5.45$ GHz, respectively. The ancilla transmon
($\omega_q/2\pi=4.79$ GHz) has two coupling pads (orange circles) that overlap with the
cavity fields. Cavity-ancilla interaction is achieved by application of a frequency-matched RF drive (green arrow) to the coupling pin near the Josephson junction (marked by X). The ancilla also serves to prepare and read out the cavity state, and is measured by its dispersive coupling to a
stripline readout resonator (orange rectangle). More details on this device can be found in
Supplementary Note 1 and Ref.~[20].
}
\end{Figure}

\begin{figure*}
\centering
\includegraphics[scale=1]{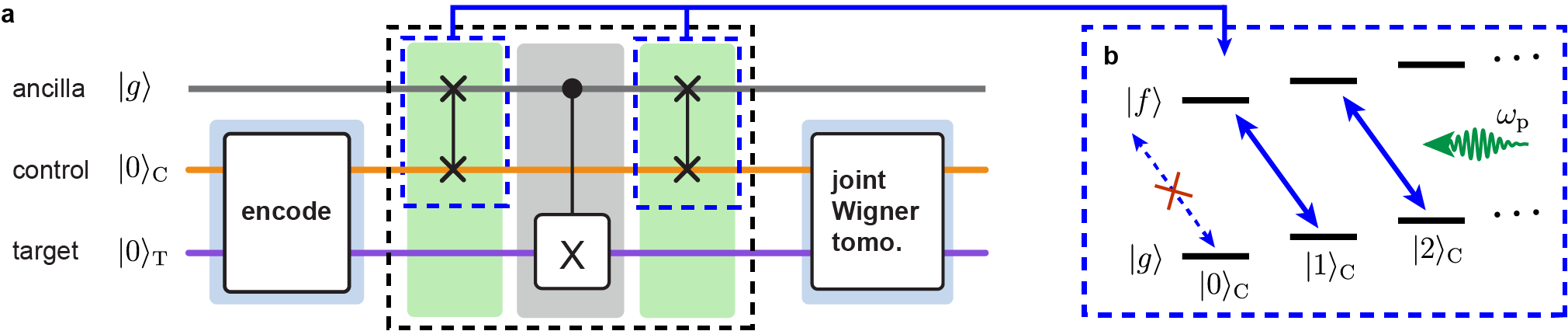}
\caption{\textbf{Protocol of the entangling gate. a,} The sequence starts with preparation of the
desired initial two-cavity state, while leaving the ancilla transmon in the ground state. The
cavity-cavity CNOT gate (dashed black rectangle) consists of two entangling gates between the
control cavity and the ancilla (dashed blue rectangles), interleaved by a CNOT gate between the
ancilla and the target, implemented by a conditional $\pi/2$ phase-space rotation of the target
cavity. The joint Wigner distribution of the final two-cavity state is measured using a method
similar to Ref.~[20]. \textbf{b,} Schematic level diagram illustrating the RF-driven control-ancilla sideband transition. Through the absorption of a single drive photon (in green) and a single
control photon, the ancilla is doubly excited from $|g\rangle$ to $|f\rangle$ (solid blue arrows).
However, when the control cavity is in vacuum, the absence of a control photon prevents the
ancilla from being excited to $|f\rangle$ (dashed blue arrow).}
\end{figure*}

\section*{Results}

\textbf{System description.} The experimental system used for implementing the entangling gate is
depicted in Fig. 1.
The multiphoton qubits are encoded in two high-Q ($T_{_1}\sim$ 0.002 s) superconducting coaxial
stub cavities. Several different multiphoton encodings are compatible with the entangling gate (see Supplementary
Note 7). Here, we choose a basis of even-parity Fock states \begin{equation}
|0_\textrm{L}\rangle_\textrm{C} = |0\rangle_\textrm{C}, \qquad |1_\textrm{L}\rangle_\textrm{C}=
|2\rangle_\textrm{C}
\end{equation}
for the control cavity, and Schr\"odinger kitten states\cite{michael_new_2016}
\begin{align}
|0_\textrm{L}/1_\textrm{L}\rangle_\textrm{T}&= \frac{1}{\sqrt{2}}\left(\frac{|0\rangle_\textrm{T}+
|4\rangle_\textrm{T}}{\sqrt{2}}\pm|2\rangle_\textrm{T}\right)
\end{align}
for the target cavity (henceforth omitting normalization). These encodings can allow error
detection of a photon loss event in both cavities, as well as error correction in the target
cavity.

The operation of the gate relies on two types of nonlinear interaction between the cavities and
the ancilla, enabled by the ancilla's Josephson junction. The first is the naturally occurring
dispersive interaction, which can be understood as a rotation of the cavity phase space
conditioned on the ancilla state. Here, we consider the ancilla ground and second excited states
$|g\rangle$ and $|f\rangle$ only, since the first excited state $|e\rangle$ is ideally not
populated during the gate operation. In this case the Hamiltonian is
\begin{equation}
\hat{H}_{\text{disp}}/\hbar=-\widetilde{\chi}_{_\textrm{T}}\hat{a}_{_\textrm{T}}^\dag\hat{a}_{_\textrm{T}}|f\rangle\langle
f|-\widetilde{\chi}_{_\textrm{C}}\hat{a}_{_\textrm{C}}^\dag\hat{a}_{_\textrm{C}}|f\rangle\langle f
|,
\end{equation}
where $\hat{a}_{_{\textrm{C(T)}}}$ is the control (target) annihilation operator. As a result of
this interaction, the target (control) cavity phase space rotates at
$\widetilde{\chi}_{_{\textrm{C(T)}}}/2\pi=1.9$ MHz (3.3 MHz) when the ancilla is in $|f\rangle$,
but remains unchanged when the ancilla is in $|g\rangle$.

We can also drive a sideband interaction between the control cavity and the ancilla using an RF
tone that satisfies the frequency matching condition
$\omega_{_\textrm{p}}=\omega_{gf}-\omega_{_\textrm{C}} -
(n_{_\textrm{C}}-1)\widetilde{\chi}_{_\textrm{C}}$, with $\omega_{gf}/2\pi=9.46$ GHz the ancilla
transition frequency between $|g\rangle$ and $|f\rangle$ (Fig. 2b), and $n_{_\textrm{C}}$ the
number of control photons (we discuss the effect of the target photon number $n_{_\textrm{T}}$ in
Supplementary Note 4). This interaction, described by the Hamiltonian
\begin{equation}
\hat{H}_{\text{sb}}/\hbar=\frac{\Omega_{_\textrm{C}}(t)}{2}\left(\hat{a}_{_\textrm{C}}|f\rangle\langle
g | +\hat{a}^\dag_{_\textrm{C}}|g\rangle\langle f |\right),
\end{equation}
leads to sideband oscillations\cite{leek_using_2009} between the states
$|n_{_\textrm{C}},g\rangle$ and
$|n_{_\textrm{C}}-1,f\rangle$\cite{zeytinoglu microwave-induced_2015,pechal_microwave-controlled_2014,gasparinetti_measurement_2016}.
By strongly driving this transition we obtain an oscillation rate of
$\sqrt{n_{_\textrm{C}}}\Omega_{_\textrm{C}}/2\pi=11.2$ MHz with $n_{_\textrm{C}}=2$, close to the
theoretical prediction (see Supplementary Note 3). However, for $n_{_\textrm{C}}=0$ the pump
does not drive sideband oscillations, and the ancilla remains in its ground state (Fig. 2b).

\textbf{Gate protocol.} The basic mechanism behind the gate is to make the cavities interact
sequentially with the ancilla, enabling an effective nonlinear interaction between the cavities
without requiring a significant direct cavity-cavity coupling. This method is similar to the one
used in a recent experiment on a gate between single optical
photons\cite{hacker_photon-photon_2016}. We start by preparing the desired initial state using
optimal control pulses on the ancilla and on the
cavities\cite{krastanov_universal_2015,heeres_cavity_2015,heeres_implementing_2016}, after
ensuring that the ancilla is initialized in $|g\rangle$. The gate sequence itself is then
performed in three steps (Fig. 2a). First, we apply the sideband drive for
$\pi/(\sqrt{2}\Omega_{_\textrm{C}})= 45$ ns, exciting the ancilla to $|f\rangle$ conditioned on
the control being in $|1_\textrm{L}\rangle_\textrm{C}$. We then turn off the drive for 100 ns
(approximately $\pi/2\widetilde{\chi}_{_\textrm{T}}$, see Supplementary Note 4), during which
the ancilla dispersively interacts with the target cavity. This flips
$|0_\textrm{L}\rangle_\textrm{T}$ into $|1_\textrm{L}\rangle_\textrm{T}$ and vice versa,
conditioned on the ancilla being in $|f\rangle$. We then apply the sideband drive a second
time to disentangle the ancilla from the cavities, thereby effectively achieving a CNOT gate
between the two cavities after a total gate time of $t_\textrm{g}\sim 190$ ns.
Finally, we use the ancilla to perform joint Wigner tomography on the two-cavity state
\cite{wang_schrodinger_2016,lutterbach_method_1997}, from which we can reconstruct the density
matrix (see Supplementary Note 6).

\textbf{Gate characterization. }
The hallmark of a CNOT gate is its ability to entangle two initially separable systems. As a
demonstration of this capability, we apply the gate to
$|\psi_{\textrm{in}}\rangle=(|0_\textrm{L}\rangle_\textrm{C}+|1_\textrm{L}\rangle_\textrm{C})\otimes|0_\textrm{L}\rangle_\textrm{T}$
(Fig. 3a). Ideally, this should result in a logical Bell state
$|\psi_\textrm{ideal}\rangle=|0_\textrm{L}\rangle_\textrm{C}|0_\textrm{L}\rangle_\textrm{T}+|1_\textrm{L}\rangle_\textrm{C}|1_\textrm{L}\rangle_\textrm{T}$.
By reconstructing the output density matrix $\rho_{\textrm{meas}}$ (Fig. 3b), we deduce a state
fidelity of $F_{\textrm{Bell}}\equiv\langle\psi_{
\textrm{ideal}}|\rho_{\textrm{meas}}|\psi_\textrm{ideal}\rangle=\left(90\pm 2\right)\%$. This is
within the measurement uncertainty of the input state fidelity  $F_{\textrm{in}}=\left(92\pm
2\right)\%$. Therefore, we conclude that the effect of nonidealities in the gate operation on the
Bell state fidelity is obscured by imperfections in state preparation and measurement (see Supplementary Note 5).

To fully characterize the CNOT gate, we next perform quantum process
tomography\cite{chuang_prescription_1997} (QPT). We achieve this by applying the gate to sixteen
logical input states that together span the entire code space. By performing quantum state
tomography on the resulting output states we can reconstruct the quantum process
$\epsilon(\rho_{\rm{in}})$, which captures the action of the gate on an arbitrary input state
$\rho_{\rm{in}}$.
The result can be expanded in a basis of two-qubit generalized Pauli operators $E_i$ on the code
space as
$\epsilon(\rho_{\rm{in}})=\sum_{m,n=0}^{15}{\chi_{m,n}E_m\rho_{\rm{in}}E_n}$,
where $\chi$ is the process matrix. Using the measured $\chi$ (Fig. 4a), we determine a process
fidelity of $\textrm{F}_\textrm{CNOT} \equiv \textrm{Tr}\left\{ \chi_\textrm{ideal} \chi
\right\}=\left(89\pm 2\right)\%$. We can estimate the effect of nonideal state preparation and
measurement by performing QPT on the process consisting of encoding and measurement only, yielding
a fidelity with the identity operator of
$\textrm{F}_{\textrm{identity}}=\left(92\pm 2\right)\%$.

\begin{Figure}
\centering
\includegraphics[scale=1]{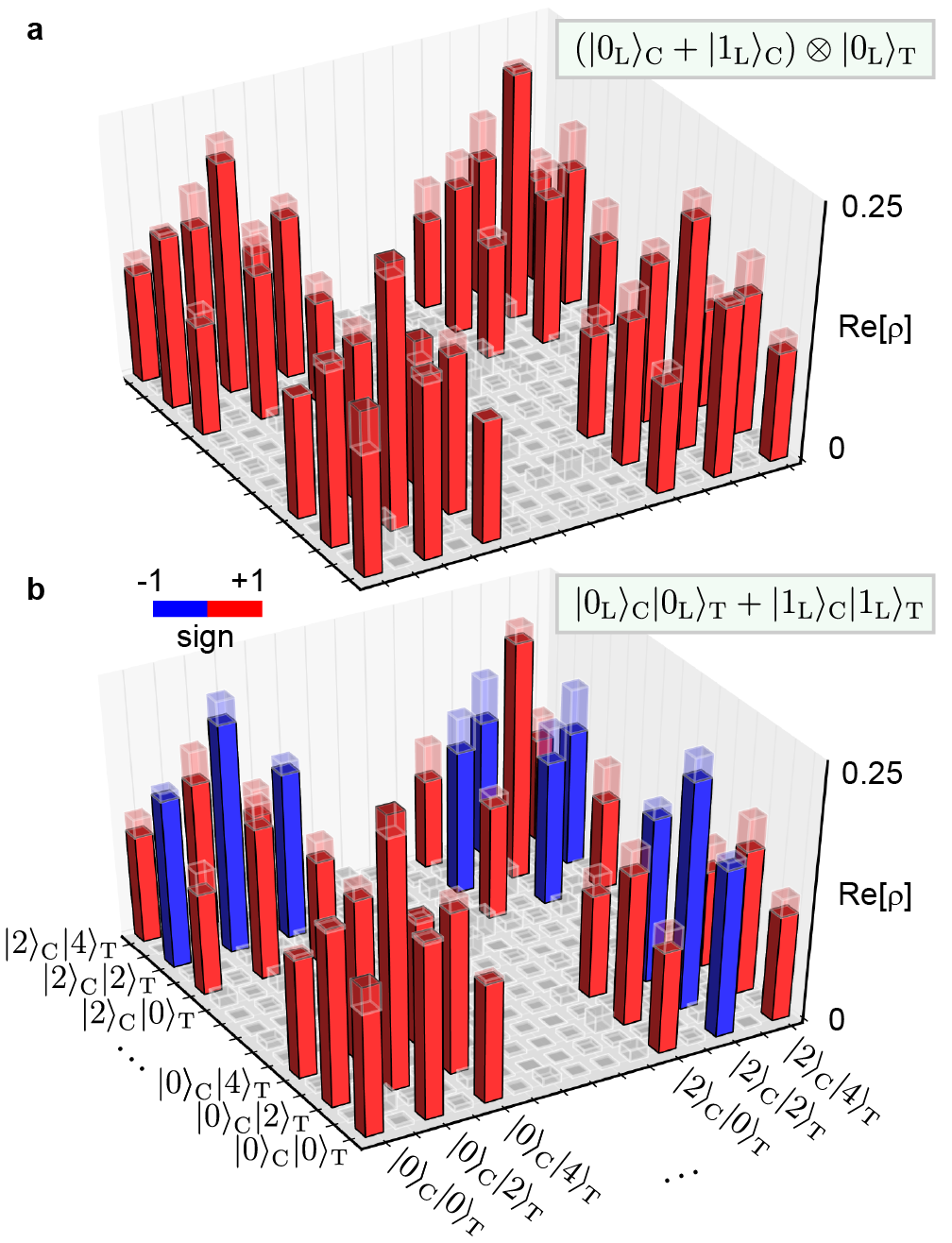}
\captionof{figure}{\textbf{Generation of a multiphoton Bell state.} Reconstructed density matrices
(solid bars) of \textbf{a,} the initial separable two-cavity state
$(|0\rangle_\textrm{C}+|2\rangle_\textrm{C})\otimes(\frac{|0\rangle_\textrm{T}+
|4\rangle_\textrm{T}}{\sqrt{2}}+|2\rangle_\textrm{T})$ (ideal shown by transparent bars) and
\textbf{b,} the output state after application of the CNOT gate, turning the kitten state into
($\frac{|0\rangle_\textrm{T}+ |4\rangle_\textrm{T}}{\sqrt{2}}-|2\rangle_\textrm{T}$), provided the
control state is $|2\rangle_\textrm{C}$. We reconstruct the density matrices assuming a Hilbert
space spanned by the Fock states $|n\rangle_\textrm{C}|m\rangle_\textrm{T}$ with $n<3$ and $m<5$
after confirming the absence of population at higher levels. Components of the density matrices
below 0.05 are colored in gray for clarity. The imaginary parts are small as well, and are shown
in Supplementary Note 8 for completeness.}
\end{Figure}

To more accurately determine the performance of the gate and highlight specific error mechanisms,
we apply it repeatedly to various input states (Fig. 4b). We then measure how the state fidelity
decreases with the number of gate applications.
A first observation is that no appreciable degradation in state fidelity occurs when the control
qubit is in $|0_\textrm{L}\rangle_\textrm{C}$. Indeed, the control cavity contains no photons in
this case, and as a result the ancilla remains in its ground state at all times. When the initial
two-cavity state is $|1_\textrm{L}\rangle_\textrm{C}|X_\textrm{L}^-\rangle_\textrm{T}$
(introducing $|X_\textrm{L}^{\pm}\rangle \equiv (|0_\textrm{L}\rangle \pm |1_\textrm{L}\rangle)$
and $|Y_\textrm{L}^{\pm}\rangle \equiv (|0_\textrm{L}\rangle \pm i|1_\textrm{L}\rangle)$),
corresponding to $|2\rangle_\textrm{C}|2\rangle_\textrm{T}$ in the Fock-state basis, the ancilla
does get excited to the $|f\rangle$-state, and we measure a small decay in fidelity of
$(0.6 \pm 0.3) \%$ per gate application. This is consistent with the ancilla decay time from
$|f\rangle$ to $|e\rangle$, measured to be 40 $\mu$s. While the qubit is irreversibly lost when a
decay occurs, the final ancilla state is outside the code space, and therefore this is a
detectable error. If the control qubit is initially in a superposition state, the first sideband
pump pulse will entangle the control cavity with the ancilla, making the state prone to both
ancilla decay and dephasing ($T_2^f=17$ $\mu$s). For example, for
$|X_\textrm{L}^+\rangle_{_\textrm{C}}|X_\textrm{L}^-\rangle_{_\textrm{T}}$, we measure a decay in
fidelity of $(0.9 \pm 0.2)\%$ per gate.
When the target cavity state is not rotationally invariant (i.e. not a Fock state), we observe
larger decay rates ($(2.0 \pm 0.3)\%$ for
$|1_\textrm{L}\rangle_\textrm{C}|0_\textrm{L}/1_\textrm{L}\rangle_\textrm{T}$, and $(2.1 \pm
0.2)\%$ for $|Y_\textrm{L}^+\rangle_{_\textrm{C}}|Y_\textrm{L}^+\rangle_{_\textrm{T}}$). Possible
mechanisms for these increased decay rates are discussed in Supplementary Note 4.
While an accurate determination of the gate fidelity would require randomized
benchmarking\cite{magesan_efficient_2012}, the data presented in Fig. 4b is sufficient to infer an
average degradation in state fidelity of approximately 1\% per gate application, close to the
$\sim 0.5\%$ limit set by ancilla decoherence.

\begin{Figure}
\centering
\includegraphics[scale=1]{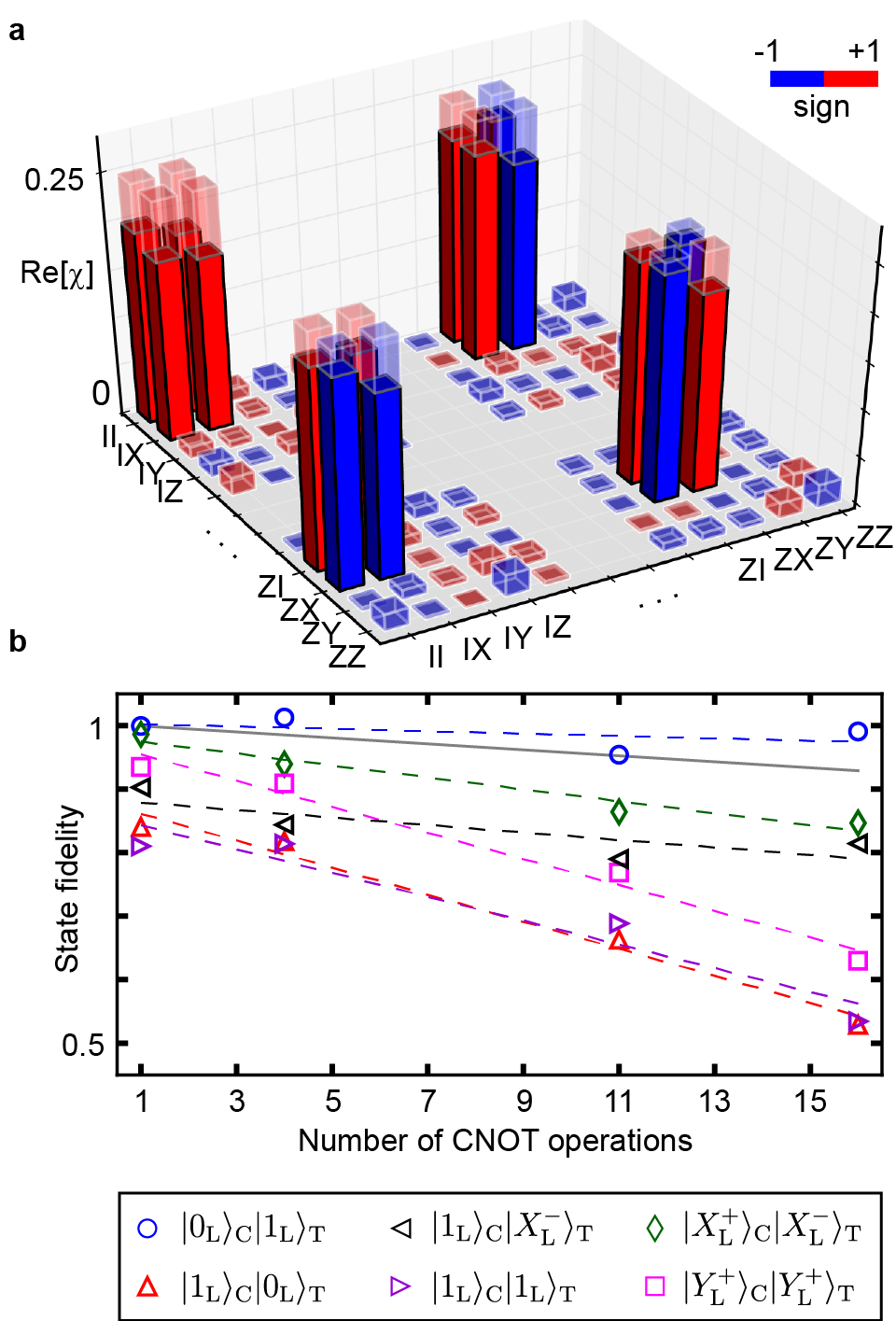}
\captionof{figure}{\textbf{Characterization of the controlled NOT gate. a,} Quantum process
tomography. The solid (transparent) bars represent the measured (ideal) elements of the process
matrix $\chi$. The corresponding process fidelity is  $\textrm{F}_\textrm{CNOT}=\left(89\pm
2\right)\%$. For clarity, only the corners
of the process matrix are presented. The full $\chi$-matrix is shown in Supplementary
Note 8 for completeness. \textbf{b,} State fidelity under repeated gate applications for
various input states, chosen to highlight different error mechanisms of the gate (the dashed lines
are linear fits). The solid gray line depicts the simulated average slope of state fidelity
imposed by ancilla decoherence. The state fidelities are calibrated by the value measured for the
vacuum state (see Supplementary Note 6). The standard errors are derived from bootstrapping, and are equal in size to the symbols.}
\end{Figure}

An important figure of merit for an entangling gate is the ability to turn off the interaction, to
prevent unwanted entanglement between the cavities. In practice, the cross-Kerr interaction
between the cavities, described by the Hamiltonian
$\hat{H}_{_\textrm{CT}}/\hbar=\chi_{_\textrm{CT}} \hat{a}^{\dag}_{_\textrm{C}}
\hat{a}_{_\textrm{C}} \hat{a}^{\dag}_{_\textrm{T}} a_{_\textrm{T}}$, induces entanglement even
when the gate operation is not applied.
To measure the interaction rate $\chi_{_\textrm{CT}}$, we prepare a separable two-cavity state in
a code space spanned by vacuum and the single-photon Fock state (see Supplementary Note 7),
and perform state tomography after variable delay times.
When extracting the concurrence\cite{wootters_entanglement_1998} of the measured density matrices
(Fig. 5), we observe first an increase, then a subsequent decrease, of the entanglement between
the cavities.

\begin{Figure}
\centering
\includegraphics[scale=1]{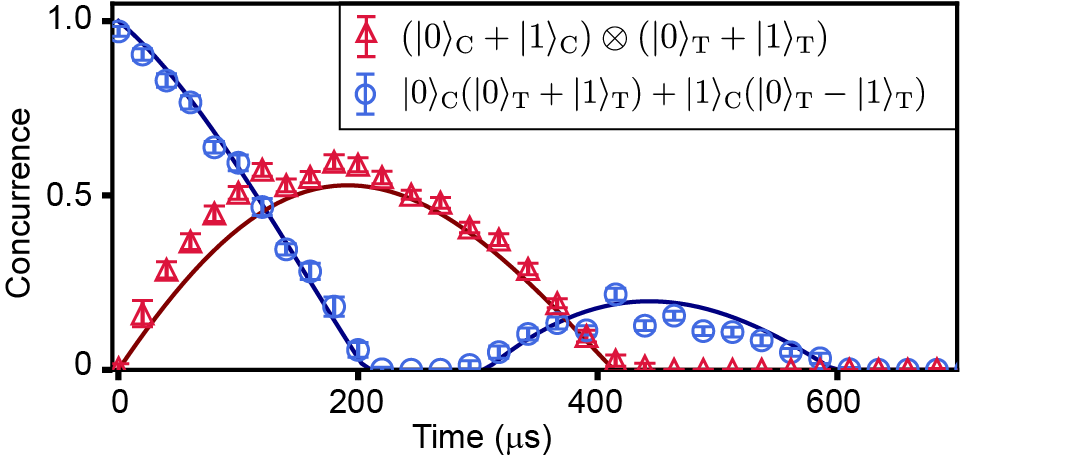}

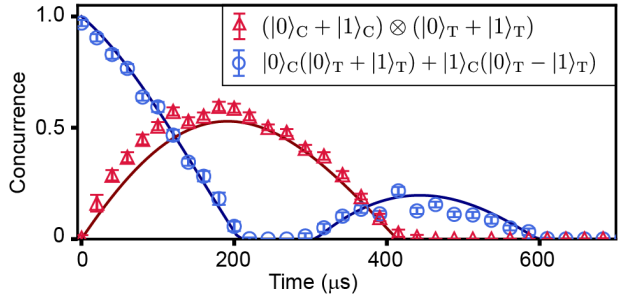
\captionof{figure}{\textbf{Undesired entanglement induced by the coupling ancilla.} Concurrence
vs. wait time for an initially separable state (red) using single-photon encoding, and for an
initial Bell state (blue) obtained by applying the CNOT gate to the separable state. The presence of
the cross-Kerr interaction between the two cavities is responsible for the observed oscillatory
behavior, whereas dephasing due to thermal excitations in the ancilla results in a gradual decay
of the entanglement. By fitting simulations (solid curves) to the measured data, we
determine a cross-Kerr interaction rate of $\chi_{_\textrm{CT}}/2\pi$=2 kHz. Error bars indicate the standard error derived from bootstrapping.}
\end{Figure}
In a similar vein, when starting with a Bell state, the cross-Kerr interaction first disentangles,
and then re-entangles the two cavities. The cavity dephasing times of $\sim 500$ $\mu$s lead to a
gradual overall loss of entanglement in both cases. From the measured curves, we infer a
cross-Kerr interaction rate of $\chi_{_\textrm{CT}}/2\pi$= 2 kHz. However, the residual
entanglement rate for the multiphoton encoding is increased to
$\Omega_{\textrm{res}}=n_{_\textrm{C}} \bar{n}_{_\textrm{T}}\chi_{_\textrm{CT}}=$ $2\pi\times$8
kHz, where $\bar{n}_\textrm{T}=2$ is the average number of photons in the target cavity. We can
therefore infer the on/off ratio of the entangling rate, defined by the ratio of the times to
generate maximal entanglement without and with gate application, to be
$\pi/(\Omega_{\textrm{res}}t_\textrm{g})\sim$ 300.

\section*{Discussion}

In conclusion, we have realized a high-fidelity entangling gate between multiphoton states encoded
in two cavities.  Together with single-qubit gates\cite{heeres_implementing_2016}, this provides a
universal gate set on encoded qubits that can be actively protected\cite{ofek_extending_2016}
against single-photon loss. The gate relies on correct operation of the control-ancilla sideband
drive, restricting the choice of control encodings. In fact, the encoding used in this
demonstration, as well as a variety of similar encodings compatible with the CNOT gate, provides
full error-correctability for the target cavity, but only detectability of a photon loss error in
the control cavity. However, a generalization of the kitten code exists which could potentially
allow for identical error-correctable encodings in both cavities (see Supplementary Note 7).
An important criterion of a gate on error-corrected logical qubits is whether errors before or
during the gate operation can be detected or corrected. Using our scheme, ancilla or cavity decay
events can be detected since they lead to a final state outside the code space. However, the gate
fidelity is ultimately limited by ancilla dephasing, and with the current encoding the control
cavity is subject to uncorrectable no-jump evolution\cite{michael_new_2016}. These remaining
imperfections need to be addressed in future fault-tolerant gate implementations. The demonstrated
gate is especially useful for practical applications that are limited by decoherence processes or
spurious interactions during long idle times. In particular, it establishes the potential of
multicavity registers for distributed quantum computing, combining long-lived storage qubits with
high-fidelity local operations \cite{jiang_distributed_2007,nickerson_topological_2013}.

\section*{Acknowledgements}
We thank K. Sliwa, M.J. Hatridge and A. Narla for providing the Josephson Parametric Converter
(JPC), K. Chou and J.Z. Blumoff for helpful discussions, and N. Ofek for providing the logic for
the field programmable gate array (FPGA) used in the control of this experiment. This research was
supported by the U.S. Army Research Office (W911NF-14-1-0011). Y.Y.G. was supported by an A*STAR
NSS Fellowship; P.R. by the U.S. Air Force Office of Scientific Research (FA9550-15-1-0015);
C.J.A. by an NSF Graduate Research Fellowship (DGE-
1122492); S.M.G. by the National Science Foundation (DMR-1609326); L.J. by the Alfred P. Sloan
Foundation and the Packard Foundation. Facilities use was supported by the Yale Institute for
Nanoscience and Quantum Engineering (YINQE), the Yale SEAS cleanroom, and the National Science
Foundation (MRSECDMR-1119826).

\section*{Author contributions}
S.R. and Y.Y.G. carried out measurements and data analysis. Devices were fabricated by C.W. and
Y.Y.G. Experimental contributions were provided by C.W., P.R., C.J.A. and L.F. Theoretical
contributions were provided by L.J. and M.M. The experiment was designed by S.R., Y.Y.G., C.W. and
R.J.S. The manuscript was written by S.R., Y.Y.G. and R.J.S. with feedback from all authors.
S.M.G, M.H.D. and R.J.S. supervised the project.

\newpage

\renewcommand{\arraystretch}{1.5}
\newcommand{\C}{\textrm{C}}
\newcommand{\T}{\textrm{T}}
\newcommand{\LL}{\textrm{L}}
\newcommand{\ro}{\textrm{r}}

\renewcommand{\figurename}{Supplementary Figure}
\renewcommand{\tablename}{Supplementary Table}
\setcounter{figure}{0}

\def\refname{Supplementary References}

\section*{Supplementary Information}

\subsection*{Supplementary Note 1: Experimental device}

\begin{Figure}
\centering
\includegraphics[scale=1]{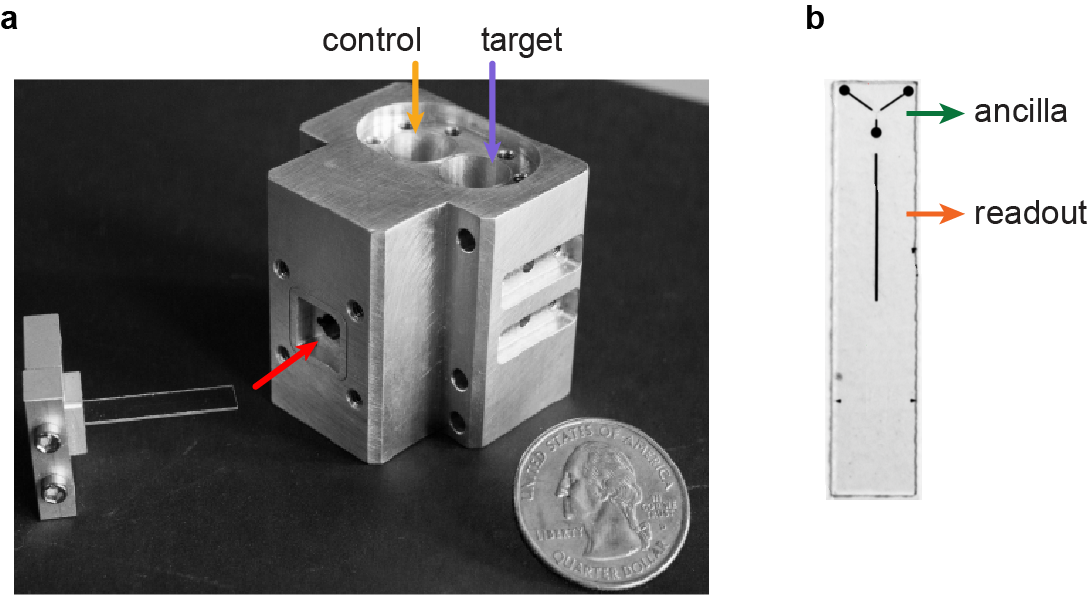}
\captionof{figure}{\textbf{Overview of the experimental device. a,} Photograph of the aluminum package housing the control and target cavities. The sapphire chip containing an ancilla transmon and a stripline readout resonator is clamped by an aluminum holder and inserted into a designated tunnel in the package. \textbf{b,} Micrograph of the sapphire chip.  }
\end{Figure}

The experimental device consists of a 3D structure made from high-purity aluminum, containing two high-Q coaxial stub cavities\cite{reagor_reaching_2013b}, as well as a tunnel in which we insert a sapphire chip containing a fixed-frequency ancilla transmon and a stripline readout resonator\cite{axline_architecture_2016b} (Supplementary Figure 1a). The fundamental modes of the superconducting cavities are used for encoding the control and target qubits\cite{reagor_quantum_2016b}. The ancilla (with anharmonicity of $2\pi\times$117 MHz) has three antenna pads providing coupling to both cavities and to the readout resonator (Supplementary Figure 1b). The ancilla and the cavities are each undercoupled to separate pins through which they are driven by optimal control pulses\cite{heeres_implementing_2016b} to prepare the desired initial states. The ancilla coupling pin is also used for applying the sideband pump tone and for driving the readout resonator. In addition, the readout resonator is overcoupled to a pin that transmits the readout signal to a Josephson parametric converter (JPC), followed by a high electron mobility transistor (HEMT) amplifier, allowing single-shot readout of the ancilla state.

 \subsection*{Supplementary Note 2: System Hamiltonian}

The four modes of the device involved in the experiment can be approximately described up to fourth order in their fields by the Hamiltonian
\begin{flalign}
\hat{H}_0/\hbar
&= \omega_{_\C}\hat{a}_{_\C}^\dag\hat{a}_{_\C}+\omega_{_\T}\hat{a}_{_\T}^\dag\hat{a}_{_\T}
  +\omega_{_\ro}\hat{a}_{_\ro}^\dag\hat{a}_{_\ro}+\omega_{ge}|e\rangle\langle e| +\omega_{gf}|f\rangle\langle f| &\nonumber\\
 -&\widetilde{\chi}_{_\C}^e\hat{a}_{_\C}^\dag\hat{a}_{_\C}|e\rangle\langle e|
  -\widetilde{\chi}_{_\T}^e\hat{a}_{_\T}^\dag\hat{a}_{_\T}|e\rangle\langle e|
  -\widetilde{\chi}_{_\ro}^e\hat{a}_{_\ro}^\dag\hat{a}_{_\ro}|e\rangle\langle e| & \nonumber\\
 -&\widetilde{\chi}_{_\C}\hat{a}_{_\C}^\dag\hat{a}_{_\C}|f\rangle\langle f|
  -\widetilde{\chi}_{_\T}\hat{a}_{_\T}^\dag\hat{a}_{_\T}|f\rangle\langle f|
  -\widetilde{\chi}_{_\ro}\hat{a}_{_\ro}^\dag\hat{a}_{_\ro}|f\rangle\langle f| &  \\
 -&\chi_{_{\C\T}}\hat{a}_{_\C}^\dag\hat{a}_{_\C}\hat{a}_{_\T}^\dag\hat{a}_{_\T}
  -\chi_{_{\C\ro}}\hat{a}_{_\C}^\dag\hat{a}_{_\C}\hat{a}_{_\ro}^\dag\hat{a}_{_\ro}
  -\chi_{_{\T\ro}}\hat{a}_{_\T}^\dag\hat{a}_{_\T}\hat{a}_{_\ro}^\dag\hat{a}_{_\ro}  & \nonumber\\
 -&\frac{\chi_{_{\C\C}}}{2}\hat{a}_{_\C}^\dag\hat{a}_{_\C}^\dag\hat{a}_{_\C}\hat{a}_{_\C}
  -\frac{\chi_{_{\T\T}}}{2}\hat{a}_{_\T}^\dag\hat{a}_{_\T}^\dag\hat{a}_{_\T}\hat{a}_{_\T}
  -\frac{\chi_{_{\ro\ro}}}{2}\hat{a}_{_\ro}^\dag\hat{a}_{_\ro}^\dag\hat{a}_{_\ro}\hat{a}_{_\ro},\nonumber&
\end{flalign}
where the readout mode is denoted by `$\ro$'.  The parameters in this Hamiltonian are specified in Supplementary Table 1, and the coherence properties of the modes are described in Supplementary Table 2. The first row in Supplementary Equation 1 describes the transition frequencies of the modes, and the second and third rows contain the dispersive interaction terms of the ancilla $|e\rangle$ and $|f\rangle$-states with the readout resonator and the cavities, which are all in the few megahertz range. The final two rows describe the cross-Kerr interaction terms between the readout and cavity modes, as well as their self-Kerr rates, which are all in the few kilohertz range. The only term in this free-evolution Hamiltonian that is explicitly used in the CNOT gate protocol is the dispersive interaction between the target cavity and the ancilla in the $|f\rangle$-state at a rate $\widetilde{\chi}_{_\T}$, which also determines the ultimate speed limit of the gate.

\begin{center}
 \captionof{table}{Parameters of the full system Hamiltonian.}
 \begin{tabular}{c c |c c}
 \hline\hline\\[-3ex]
 	 Term ($/2\pi$)	& 	Measured  &	 Term ($/2\pi$)	& 	Measured\\  [-1.5ex]
 	      & (Predicted)	&  & 	(Predicted)\\
  \hline\hline
  $\omega_{_\C}$& \,4.22 GHz&$\omega_{ge}$& 4.79 GHz \\
  $\omega_{_\T}$& \,5.45 GHz&$\omega_{gf}$& 9.46 GHz\\
  $\omega_{_\ro}$& \,7.70 GHz&           & \\
  \hline
  $\widetilde{\chi}_{_\C}^e$& \,1.02 MHz&$\widetilde{\chi}_{_\C}$& 3.3 MHz\\
  $\widetilde{\chi}_{_\T}^e$& \,1.10 MHz&$\widetilde{\chi}_{_\T}$& 1.9 MHz\\
  $\widetilde{\chi}_{_\ro}^e$& \,1.74 MHz&$\widetilde{\chi}_{_\ro}$& (3.3 MHz)\\
  \hline
  $\chi_{_{\C\T}}$& \,2 kHz&$\chi_{_{\C\C}}$& (1.6 kHz)\\
  $\chi_{_{\C\ro}}$& \,(5 kHz)&$\chi_{_{\T\T}}$& (3.4 kHz)\\
  $\chi_{_{\T\ro}}$& \,(12 kHz)&$\chi_{_{\ro\ro}}$& (7 kHz)\\
[1ex]
 \hline\\
\end{tabular}

 \captionof{table}{\textbf{Coherence properties.} Energy relaxation time ($T_1$), dephasing time ($T_2^*$), and thermal population ($P_e$) of the system components. The ancilla and cavity states are measured before each run of the experiment to verify the absence of thermal excitations.}
 \begin{tabular}{c c c c} % centered columns (4 columns)
 \hline\hline\\[-4ex]
    	            			& $T_1$		& $T_2^*$  & $P_e$  \\
 \hline\\[-2ex]
 Control cavity:	            & \,$\sim$2.2 ms\,	& \,$\sim$0.5 ms 	& \,2-3\%\\
 Target cavity:                 & \,$\sim$2.0 ms\,	& \,$\sim$0.6 ms 	& \,2-3\% \\
  Readout resonator:                 & \,300 ns\,	& \,N/A 	& \,$<$0.2\% \\
 Ancilla $|e\rangle$: 	& \,60 $\mu$s\,	& \,37$\pm$5 $\mu$s 	& \, 7.5\%\\
  Ancilla $|f\rangle$: 	& \,40 $\mu$s\,	& \,17$\pm$5 $\mu$s 	& \, $\sim$0.5\%\\
 [0.5ex]
 \hline
 \\
\end{tabular}
\end{center}

 \subsection*{Supplementary Note 3: Driven cavity-ancilla sideband interaction}

The source of nonlinearity in our system is the Josephson junction of the ancilla transmon, whose Hamiltonian is\cite{leghtas_confining_2015b}
  \begin{flalign}
  \hat{H}_{\textrm{J}}&=-E_{\textrm{J}} \cos\left[\phi_q\left(\hat{q}+\hat{q}^\dag+\xi(t)+\xi^*(t)\right)\right.& \nonumber\\
 &\left.+\phi_\C\left(\hat{a}_\C+\hat{a}^\dag_\C\right)+\phi_\T\left(\hat{a}_\T+\hat{a}^\dag_\T\right)+\phi_\ro\left(\hat{a}_\ro+\hat{a}^\dag_\ro\right)\right],&
 \end{flalign}

where $E_\textrm{J}/h$ = 21 GHz is the Josephson energy, and $\hat{q}$ is the ancilla mode annihilation operator. $\phi_k$ are the normalized zero point flux fluctuations across the junction due to mode $k$, and $\xi(t)\approx\frac{g(t)}{\Delta}$ is the displacement of the ancilla mode when driven by a pump tone at a rate $g(t)$ and detuning $\Delta$.
In the limit of small flux through the junction, this Hamiltonian can be approximated by the fourth order term in the expansion of the cosine. In this limit, the Josephson junction acts as a four-wave mixing element. When the junction is driven by a pump, terms that would otherwise be non-energy conserving can be accessed. In particular, the pumped four-wave mixing interaction of interest to us is
 \begin{equation}
 \hat{H}_\text{sb}=-\frac{1}{2}E_{\textrm{J}}\phi_q^3\phi_\C\xi(t)\left(\hat{a}_\C\hat{q}^\dag\hat{q}^\dag+\hat{a}^\dag_\C\hat{q}\,\hat{q}\right),
 \end{equation}
which describes a sideband transition between the control cavity and the ancilla. Through this interaction a single pump photon is absorbed, while extracting a single photon from the control cavity and doubly exciting the ancilla (see Fig. 2b in the main text). This term can be made resonant provided the pump satisfies the frequency condition $\omega_{_\textrm{p}}=\omega_{gf}-\omega_{_\C} - (n_{_\C}-1)\widetilde{\chi}_{_\C}$, (we discuss the effect of $n_{_\T}$ in the next section).
This interaction is a natural choice for implementing the CNOT gate. This can be seen by noting that any pumped cavity-ancilla interaction needs to involve at least a single pump photon,  a single cavity photon and a single ancilla excitation. However, the four-wave mixing interaction requires a fourth additional photon. Since the contribution of the ancilla to the junction's zero point flux fluctuation ($\phi_q=0.32$) is stronger than that of the control cavity ($\phi_{\C}=0.016$) or that of the pump (assuming $|\xi(t)|<1$), a second ancilla excitation provides the fastest possible nonlinear interaction.

The fourth-order approximation of the cosine Hamiltonian is valid only for $\phi_q |\xi(t)|\ll 1$, setting a limit on how strongly we can pump the sideband interaction. When approaching this limit, the interaction strength will saturate, and higher-order spurious nonlinear processes are likely to appear as well. In order to calibrate the pump strength, we measure the Stark shift of the ancilla frequency (due to the term $-E_\textrm{J}\phi_q^4|\xi(t)|^2\hat{q}^\dag\hat{q}$). From this, we can infer a pump strength corresponding to $\phi_q \xi\sim$ 0.16 (or $\xi\sim$ 0.5). Supplementary Equation 3 then predicts an oscillation rate of $\sqrt{2}\Omega_{_\C}\sim 2\pi \times$11 MHz with two photons in the control cavity, in close agreement with the measured value.

\begin{figure*}
\centering
\includegraphics[scale=1]{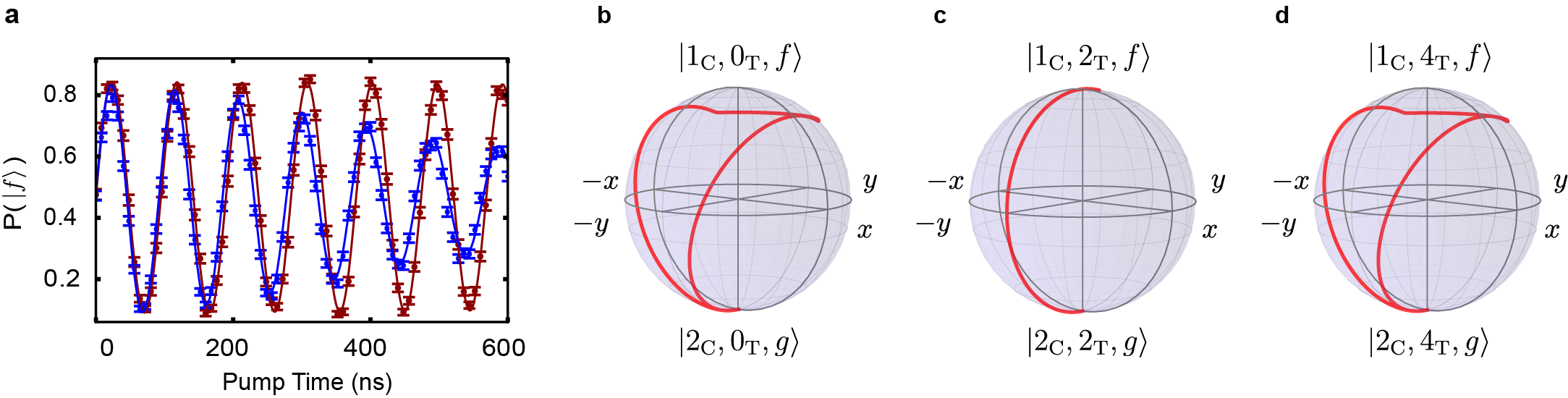}
\captionof{figure}{\textbf{Effect of the target-ancilla dispersive interaction on the sideband transitions. a,} The sideband transition is on resonance when two photons are present in the target cavity (red curve). If the target cavity contains a kitten state instead, the dispersive interaction with the ancilla results in a gradual dephasing of the sideband oscillations (blue curve). Error bars indicate the standard error of the mean. \textbf{b, c, d,}  Simulated Bloch sphere trajectories for the gate protocol (pump pulse - wait - pump pulse), with $n_{_\T}$ = 0 (\textbf{b}), 2 (\textbf{c}) and 4 (\textbf{d}). In this simulation, $\widetilde{\chi}_{_\T}/2\pi$ = 1.9 MHz and $\sqrt{2}\Omega_{_\C}/2\pi$ = 11 MHz. By choosing a 49 ns pump pulse and a 83 ns wait time, we end up in the ground state regardless of $n_{_\T}$, and acquire a phase of $\pi$ for $n_{_\T}=0$ and $n_{_\T}=4$ with respect to $n_{_\T}=2$.}
\end{figure*}

\subsection*{Supplementary Note 4: Nonidealities in the gate protocol}

In the discussion of the gate protocol in the main text, we ignored the effect of the dispersive ancilla-target interaction $\widetilde{\chi}_{_\T}\hat{a}_{_\T}^\dag\hat{a}_{_\T}|f\rangle\langle f|$ on the pumped ancilla-control sideband interaction. In reality, however, different photon numbers $n_{_\T}$ in the target cavity translate into different pump frequency matching conditions
$\omega_{_p}=\omega_{gf}-\omega_{_\C} - (n_{_\C}-1)\widetilde{\chi}_{_\C} - n_{_\T}\widetilde{\chi}_{_\T}$. If we match the pump frequency for $n_{_\T}=2$, the sideband oscillations for $|0_\T\rangle$ and $|4_\T\rangle$ will have a contrast reduced by approximately $(2\widetilde{\chi}_{_\T})^2/(\sqrt{2}\Omega_{_\C})^2\sim 10\%$. Therefore, we cannot excite the ancilla to $|f\rangle$ for all target states simultaneously (See Supplementary Figure 2a). At first sight this appears to be a limiting factor of the gate. However, complete excitation to $|f\rangle$ is not required for the CNOT gate, as long as the ancilla returns to the ground state by the end of the operation, and provided $|2_\T\rangle$ acquires a relative phase of $\pi$ with respect to $|0_\T\rangle$ and $|4_\T\rangle$. Indeed, these requirements can be met by appropriate tuning of the gate parameters. To see this, consider first the case of $|2_\C,2_\T,g\rangle$, for which the pump is on resonance. Regardless of the pump pulse duration $t_\textrm{p}$ and the wait time $t_\textrm{w}$, the area enclosed during the trajectory on the Bloch sphere composed by the two levels $|2_\C,2_\T,g\rangle$ and $|2_\C,1_\T,f\rangle$ (Supplementary Figure 2c) is zero (assuming that the initial and final pulse have opposite phases). The goal is then to make $|2_\C,0_\T,g\rangle$ and $|2_\C,4_\T,g\rangle$ acquire a total (geometric and dynamic) phase of $\pi$ during a closed trajectory on the Bloch sphere. Since the sideband transitions are detuned from resonance by an equal amount $\pm 2\widetilde{\chi}_{_\T}$ for both states, both will trace identical trajectories on the Bloch sphere, albeit in opposite directions. A closed trajectory can always be achieved by first fixing $t_\textrm{p}$, and then choosing $t_\textrm{w}$ such that the Bloch vector rotates to its mirror image with respect to the axis of rotation (see Figs. S2b-d). The second sideband pulse will then always bring the ancilla back to its ground state. Next, $t_\textrm{p}$ can be varied together with $t_\textrm{w}$ obtained by the above procedure, until a total phase of $\pi$ is acquired. Due to a nonzero ratio $\widetilde{\chi}_{_\T}/\Omega_{_\C}$, the resulting wait time and pulse duration will deviate from the simplified expressions provided in the main text $t_\textrm{w}=\pi/2\widetilde{\chi}_{_\T}=130$ ns and $t_\textrm{p}=\pi/(\sqrt{2}\Omega_{_\C})=45$ ns. Instead, a simulation gives values of $t_\textrm{w}^\text{sim}=$ 83 ns and $t_\textrm{p}^\text{sim}=$ 49 ns (Supplementary Figure 2b-d). However, in the actual experiment, $t_\textrm{w}=$100 ns and $t_\textrm{p}=$45 ns are found to satisfy the above requirements. This discrepancy is explained by a lowering of the ancilla-target dispersive interaction when the pump is switched on from $\widetilde{\chi}_{_\T}=2\pi \times 1.9$ MHz to $\widetilde{\chi}_{_\T}^\text{pump}=2\pi \times 1.4$ MHz.
For this procedure to work, it is important for both $n_{_\T}=4$ and $n_{_\T}=0$ to acquire the same total phase, which is ideally obtained by setting the pump frequency exactly on resonance for $n_{_\T}=2$. However, in the experiment, a $\sim 2\pi\times$1 MHz deviation from this frequency, or a higher-order nonlinearity of the same magnitude, may be responsible for the relatively large gate infidelity whenever the target cavity state is not rotationally symmetric, as observed in Fig. 4b of the main text. Indeed, state tomography on those states confirms that the infidelity originates mainly from a relative phase between $n_{_\T}=0$ and $n_{_\T}=4$.

Single-qubit rotations are another effect that needs to be taken into account. In the case of the control qubit, the deterministic rotation acquired due to Stark shifts and dispersive interaction with the ancilla can be annulled by fixing the phase of the second sideband pump pulse accordingly. For the target qubit, the phase acquisition due to the Stark shift is independently measured and removed in post-processing of the data.

\subsection*{Supplementary Note 5: State Preparation And Measurement errors}

As emphasized in the main text, quantum state tomography and quantum process tomography (QPT) are limited by imperfections in state preparation and measurement. An estimate of the minimal infidelity of the initial states is provided by the ratio of the control pulse duration to the ancilla dephasing time, given by 1 $\mu$s / 37 $\mu$s $\approx$ 3\%. The multiple mechanisms leading to measurement errors are discussed in detail in Ref.~[6].

\subsection*{Supplementary Note 6: Density matrix reconstruction}

For reconstructing the density matrix $\hat{\rho}_{\C\T} $ of the joint two-cavity system, we first measure its joint Wigner distribution. This is done by measuring the joint parity $\hat{P}=\text{exp}\left[{i\pi\left(\hat{a}_{_\C}^\dag\hat{a}_{_\C}+\hat{a}_{_\T}^\dag\hat{a}_{_\T}\right)}\right]$ of the cavities\cite{wang_schrodinger_2016b} after displacing them in their four-dimensional phase space:
\begin{equation}\label{eq:wig}
W_\textrm{J}\left(\beta_\C,\beta_\T\right) = \frac{4}{\pi^2}\mathrm{Tr}\left[ \hat{\rho}_{\C\T} \hat{D}_{\beta_\C}  \hat{D}_{\beta_\T}  \hat{P} \hat{D}_{\beta_\C}^\dag  \hat{D}_{\beta_\T}^\dag \right],
\end{equation}
where  $\hat{D}_{\beta}=e^{\beta \hat{a}_i^\dag - \beta^* \hat{a}_i}$ is a displacement by $\beta$ of the state of cavity $i=\C,\T$. Joint parity measurements are performed by a Ramsey interferometry measurement on the ancilla, which is subsequently read out. To compensate for imperfections in this procedure\cite{wang_schrodinger_2016b}, we calibrate the parity measurements using the value obtained for the vacuum state ($0.79 \pm 0.02$).
If we assume cutoffs $N_\C$ and $N_\T$ of the photon numbers in the control and target cavities, we can write $\hat{\rho}_{\C\T} $ as a $(N_\C N_\T)\times (N_\C N_\T)$ matrix. By measuring the joint Wigner distribution at this number of displacements or more, we can perform a maximum likelihood estimation to infer the most probable positive semi-definite Hermitian matrix $\hat{\rho}_{\C\T} $. In practice, we use $6^4$ different displacements, and reconstruct the two-cavity density matrix assuming fewer than six photons per cavity. We then confirm that up to the measurement accuracy there are at most four photons in the target cavity, and at most two photons in the control cavity for all measured states. This allows us then to reconstruct $\hat{\rho}_{\C\T}$ for this restricted 15-dimensional Hilbert space, using a now overcomplete set of data.

Since the trace of the density matrix is not constrained to unity, this method does not make the a priori assumption that the gate operation is uncorrelated with tomography errors. Instead, failures of tomography as a result of the gate operation will show up as a reduced trace, and hence a reduced state fidelity of the final density matrix.

\subsection*{Supplementary Note 7: Encodings compatible with the CNOT gate}

\subsubsection*{Single-photon encoding}
The simplest possible encoding compatible with the gate protocol presented in the main text is the single-photon encoding, with $|0\rangle_\C$($|0\rangle_\T$) and $|1\rangle_\C$($|1\rangle_\T$) as the basis states for the control (target) cavity. While this encoding offers the longest possible qubit lifetimes, it cannot be used for either error detection or error correction.
The gate protocol is identical to that of the multiphoton encoding, albeit with different timings. The presence of a single photon in the control cavity instead of two reduces the sideband oscillation rate by a factor of $\sqrt{2}$ due to the absence of bosonic enhancement. The pump tone is therefore applied for 64 ns instead of 45 ns. If the ancilla is excited to $|f\rangle$, the dispersive interaction between the target and the ancilla sets in. However, for the single-photon encoding, the rotation in the target cavity phase space required for turning   $|0\rangle_\T+|1\rangle_\T$ into  $|0\rangle_\T-|1\rangle_\T$ is $\pi$ instead of $\pi/2$, making this step about twice as long. In total, the gate time is 340 ns instead of 190 ns for the multiphoton encoding. We perform QPT for the resulting gate (Supplementary Figure 3), and measure a process fidelity of $\textrm{F}_\textrm{CNOT} =\left(95\pm 2\right)\%$. As in the case of the multiphoton encoding, state preparation and measurement cannot be isolated from the gate operation. Indeed, when performing QPT on the process consisting of encoding and readout only, we observe a similar value for the process fidelity with the expected identity operator of
$\textrm{F}_{\textrm{identity}}=\left(98\pm 1\right)\%$. We therefore conclude that the effect of gate errors on the gate fidelity is obscured by state preparation and measurement errors.
The single-photon encoding is used for the initial states in Fig. 5 of the main text. The density matrices of these states are presented in Supplementary Figure 4.

\begin{Figure}
\centering
\includegraphics[scale=1]{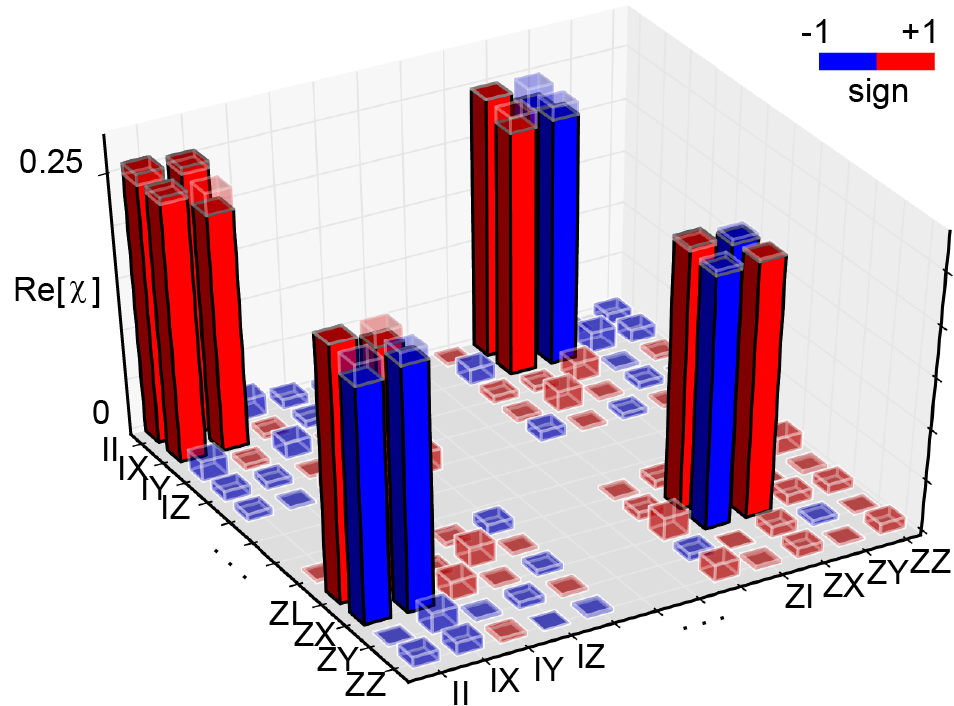}
\captionof{figure}{\textbf{Quantum process tomography of the CNOT gate with single-photon encoding.} The solid (transparent) bars represent the measured (ideal) elements of the process matrix $\chi$. The corresponding process fidelity is  $\textrm{F}_\textrm{CNOT}=\left(95\pm 2\right)\%$. For clarity, only the corners
of the process matrix are presented here.}
\end{Figure}

The single-photon encoding, as well as the multiphoton encoding used in the main text  are just two instances of a class of encodings that is compatible with the gate operation. In the control cavity, any encoding of the form $|0_\LL\rangle_\C=|0\rangle_\C,|1_\LL\rangle_\C=|n\rangle_\C$ can be used. For the target cavity, any two orthogonal states satisfying $|0_\LL\rangle_\T=e^{\pm i \hat{a}_{_\T}^\dag\hat{a}_{_\T}\theta}|1_\LL\rangle_\T$ are compatible with the gate, since these states are interchanged by a phase space rotation. However, for this class of encodings a photon loss in the control cavity can only be detected, and not corrected. This is because the control cavity collapses to $|n-1\rangle_{_\C}$ regardless of the initial state, thereby losing the stored information. Moreover, in the absence of a photon loss event, the state is gradually projected onto vacuum. This loss of information does not lead to a final state outside the code space, and is therefore an undetectable error mechanism.

\subsubsection*{Generalized kitten encoding}

The multiphoton encodings discussed above are different for both qubits, and can enable full error correction only for the target qubit.
Conventional error-correctable states such as the kitten states cannot be used for encoding the control qubit in the current scheme, since for the Fock states $|2\rangle$ and $|4\rangle$ the ratio of the sideband oscillation rates $\sqrt{n_{_\C}}\Omega_{_\C}$ is irrational.

However, we can introduce a generalized kitten encoding with basis states $|0_\LL\rangle =\frac{1}{\sqrt{2}}\left( \frac{\sqrt{3}|0\rangle+ |8\rangle}{2}+|2\rangle\right)$ and $|1_\LL\rangle = \frac{1}{\sqrt{2}}\left(\frac{\sqrt{3}|0\rangle+ |8\rangle}{2}-|2\rangle\right)$. These states have equal photon loss probability, and collapse onto orthogonal states when a single photon is lost. Therefore, this encoding satisfies the criteria for error-correctability \cite{michael_new_2016b}. In contrast to kitten states, these generalized kitten states are potentially compatible with the CNOT gate, and can be used to encode both the control and target qubits. A pump pulse that results in a full sideband transition $|2\rangle_\C|g\rangle\rightarrow |1\rangle_\C|f\rangle$ corresponds to a back and forth transition for $|8\rangle_\C|g\rangle$. As a result, only $|2_\C\rangle$ leads to an excitation of the ancilla to $|f\rangle$, and therefore to a rotation of the target cavity phase space, whereas  $\frac{1}{2}(\sqrt{3}|0_\C\rangle+ |8_\C\rangle)$ leaves the target qubit unchanged. In addition, this encoding does not entail a decrease in qubit lifetime as compared to the kitten encoding, since the average number of photons remains two. However, the presence of higher photon states requires a lower ratio $\widetilde{\chi}_{_{\T(\C)}}/\Omega_{_\C}$ than that provided by the present experimental setup, and may also increase susceptibility to higher order terms that are not taken into account in this work.

\begin{Figure}
\centering
\includegraphics[scale=1]{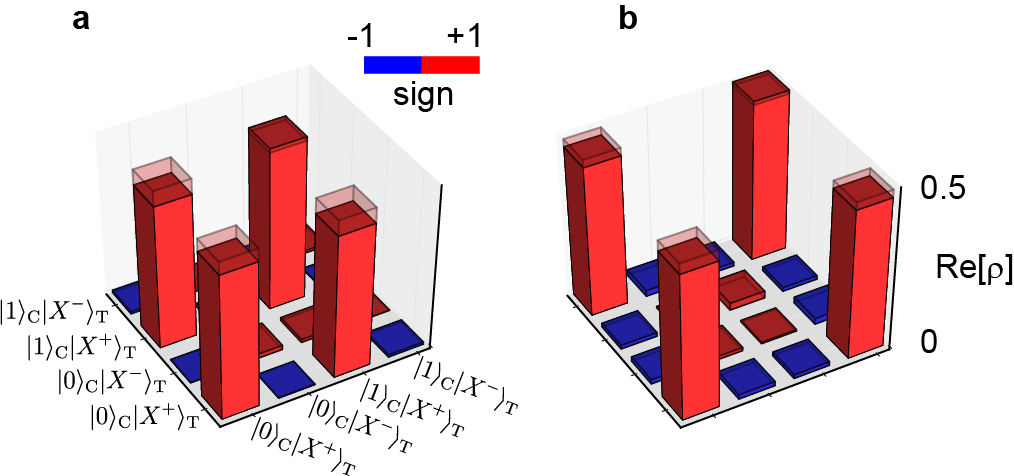}
\captionof{figure}{\textbf{Entangled state generation in the single-photon encoding.} Real parts of the reconstructed density matrices (solid bars) of \textbf{a,} the initial separable two-cavity state $(|0\rangle_\C+|1\rangle_\C)\otimes(|0\rangle_\T+|1\rangle_\T)$ (ideal shown in transparent bars), and \textbf{b,} the entangled state $|0\rangle_\C(|0\rangle_\T+|1\rangle_\T) + |1\rangle_\C(|0\rangle_\T-|1\rangle_\T)$ after application of the CNOT gate. For clarity, the target state is shown in the basis $|X^\pm\rangle_\T=(|0\rangle_\T \pm |1\rangle_\T)$.}
\end{Figure}

\subsection*{Supplementary Note 8: Additional Data}
In Supplementary Figure 5 we show the imaginary parts of the density matrices of the separable input state and the multiphoton Bell state obtained after application of the CNOT gate. The real parts are presented in Fig. 3 of the main text.

In Supplementary Figure 6, the full process matrix of the CNOT gate is presented. Only the corners of the real part are presented in Fig. 4a of the main text for clarity.

 \begin{Figure}
\includegraphics[scale=.85]{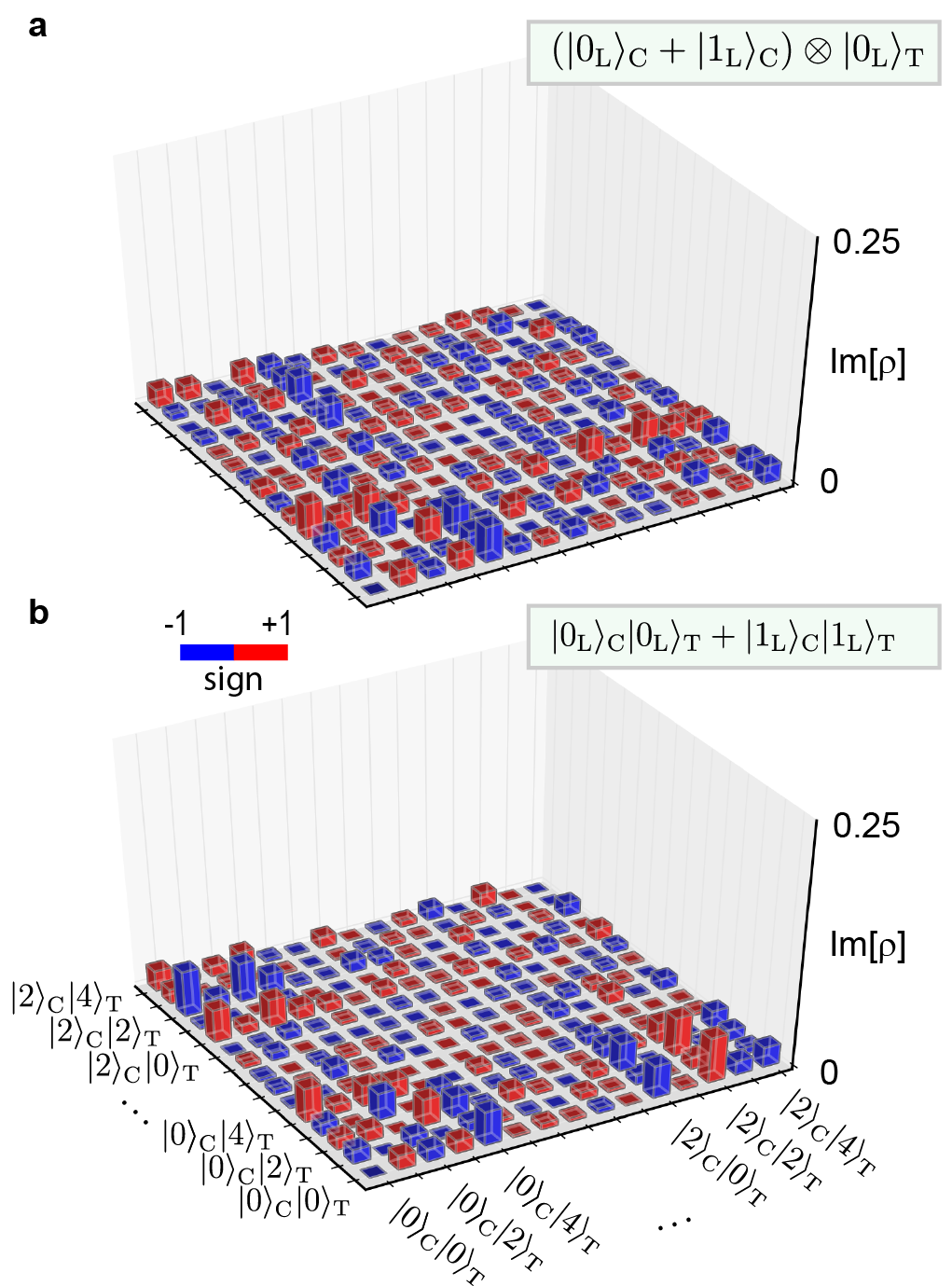}
\captionof{figure}{\textbf{Generation of a multiphoton Bell state.} Imaginary parts of the reconstructed density matrices of \textbf{a,} the initial separable two-cavity state $(|0\rangle_\C+|2\rangle_\C)\otimes(\frac{|0\rangle_\T+ |4\rangle_\T}{\sqrt{2}}+|2\rangle_\T)$ and \textbf{b,} the multiphoton Bell state after application of the CNOT gate. The real parts are shown in Fig. 3 of the main text.}
\end{Figure}

\begin{Figure}
\includegraphics[scale=.85]{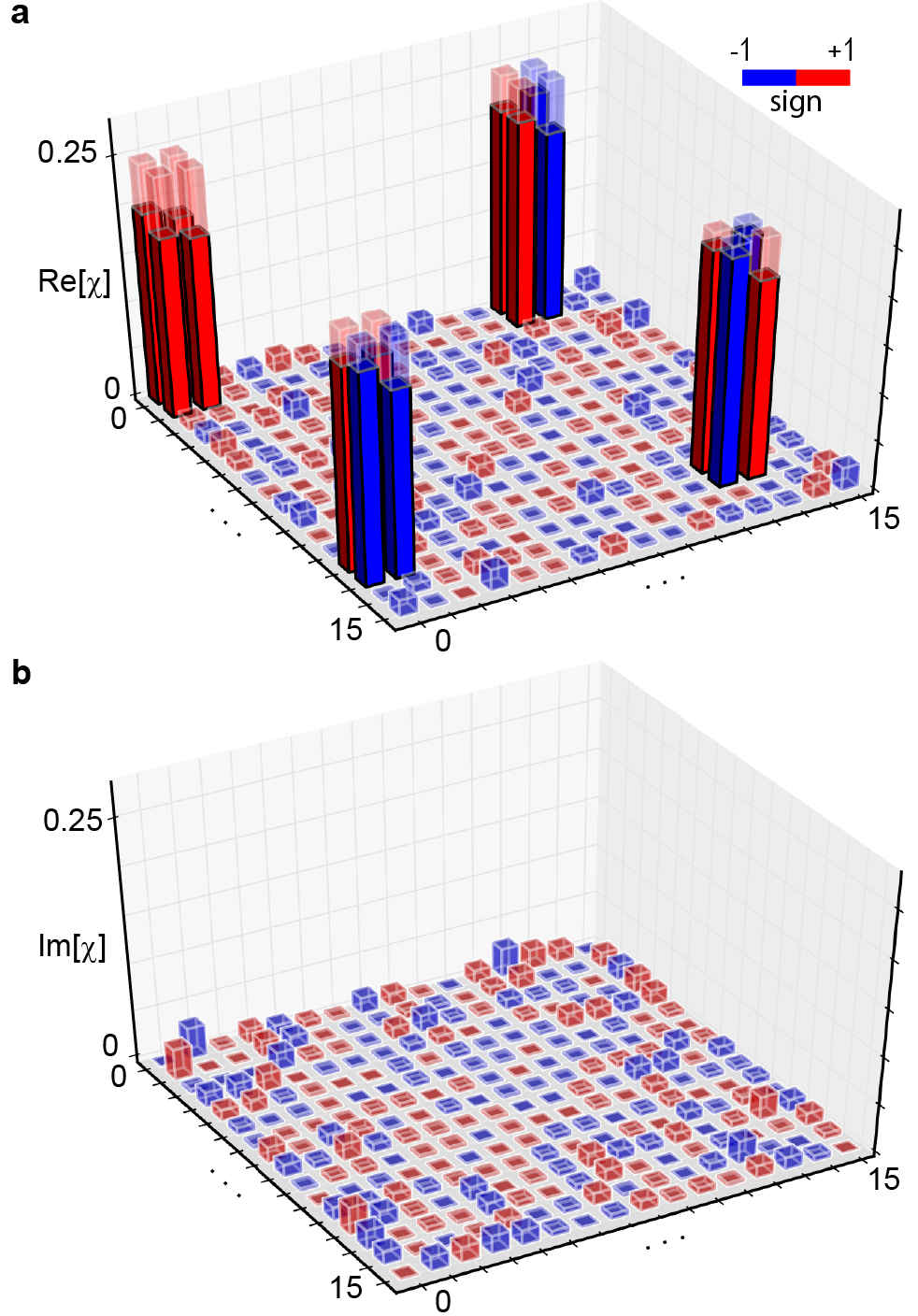}
\captionof{figure}{\textbf{Quantum process tomography. a,} Real and \textbf{b,} imaginary part of the full process matrix $\chi$ for the multiphoton encoding. The solid (transparent) bars represent the measured (ideal) elements of the process matrix. The indices correspond to the two-qubit logical Pauli operators $\left\{\text{I,X,Y,Z}\right\}^{\otimes 2}=\left\{\text{II,IX,IY,...,ZZ}\right\}$.}
\end{Figure}

\end{multicols}
\end{document}